\pdfoutput=1
\documentclass[sigconf,svgnames,table]{acmart}
\acmConference[ESEC/FSE 2017]{11th Joint Meeting of the European Software
  Engineering Conference and the ACM SIGSOFT Symposium on the Foundations of
  Software Engineering}{4--8 September, 2017}{Paderborn, Germany}

\newcommand*{\EXTENDEDVERSION}{}

\ifdefined\EXTENDEDVERSION
\renewcommand\footnotetextcopyrightpermission[1]{}
\fi

\usepackage{amsmath}
\usepackage{amssymb}
\usepackage{array, multirow}
\usepackage{comment}
\usepackage{enumitem}
\ifdefined\EXTENDEDVERSION
\else
\usepackage{flushend}
\fi
\usepackage{graphicx}
\usepackage[final]{listings}
\usepackage{pifont}
\usepackage{relsize}
\usepackage{stmaryrd}
\usepackage{syntax}
\usepackage{textcomp}
\usepackage{url}
\usepackage{xspace}

\usepackage{amsthm}

\usepackage{tikz}
\usepackage{pgfplots}
\usetikzlibrary{calc,trees,positioning,arrows,chains,shapes.geometric,%
  decorations.pathreplacing,decorations.pathmorphing,decorations.text,shapes,%
  matrix,shapes.symbols,patterns,shadows}

\newcommand{\code}[1]{\texttt{#1}}

\newcommand{\irule}[2]%
{\mkern-2mu\displaystyle\frac{#1}{\vphantom{,}#2}\mkern-2mu}

\newcommand{\prog}{{\mathcal P}}

\newcommand{\aliasing}{\Lambda}
\newcommand{\summary}{\Upsilon}
\newcommand{\tc}{\Phi}
\newcommand{\pair}[2]{\langle #1, #2 \rangle }
\newcommand{\success}{\checkmark}
\newcommand{\status}{\varphi}
\newcommand{\asrtfail}{\lightning}
\newcommand{\asmfail}{\diamondsuit}

\newcommand{\yes}{\tikz\fill[scale=0.4,color=green](0,.3) -- (.2,0) --  (.55,.55) -- (.2,.15) -- cycle;}

\newcommand{\no}{\tikz\draw[scale=0.4,color=red,line width=1.1](0.3,.7) -- (.7,0.3) (0.3,0.3) -- (.7,.7);}

\newcommand{\mv}[1]{{#1}}

\newenvironment{proofsketch}{%
  \renewcommand{\proofname}{Proof sketch}\proof}{\endproof}

\definecolor{indigo}{HTML}{283593}
\definecolor{red}{HTML}{D50000}
\definecolor{green}{HTML}{558B2F}
\definecolor{orange}{HTML}{FF9800}
\definecolor{lightgreen}{HTML}{C5E1A5}
\definecolor{lightlime}{HTML}{B2FF59}
\definecolor{lightindigo}{HTML}{9FA8DA}
\definecolor{lightpurple}{HTML}{E1BEE7}
\definecolor{lightblue}{HTML}{90CAF9}
\definecolor{lightorange}{HTML}{FFB74D}
\definecolor{light-gray}{gray}{0.6}
\definecolor{super-light-gray}{gray}{0.77}

\makeatletter
\newenvironment{btHighlight}[1][]
{\begingroup\tikzset{bt@Highlight@par/.style={#1}}\begin{lrbox}{\@tempboxa}}
{\end{lrbox}\bt@HL@box[bt@Highlight@par]{\@tempboxa}\endgroup}

\newcommand\btHL[1][]{%
  \begin{btHighlight}[#1]\bgroup\aftergroup\bt@HL@endenv%
}
\def\bt@HL@endenv{%
  \end{btHighlight}%
  \egroup
}
\newcommand{\bt@HL@box}[2][]{%
  \tikz[#1]{%
    \pgfpathrectangle{\pgfpoint{0.3pt}{0pt}}{\pgfpoint{\wd #2}{\ht #2}}%
    \pgfusepath{use as bounding box}%
    \node[anchor=base west,fill=lightorange,outer sep=0pt,inner xsep=0.3pt,inner ysep=0pt,minimum height=\ht\strutbox+0.3pt,#1]{\raisebox{0.3pt}{\strut}\strut\usebox{#2}};
  }%
}
\makeatother

\lstdefinestyle{basic}{%
  morekeywords     = [2]{assert, assume},%
  keywordstyle     = \bfseries\color{DarkBlue},%
  commentstyle     = \color{Crimson},%
  basicstyle       = \small\ttfamily,%
  columns          = [c]fixed,%
  aboveskip        = 0mm,%
  belowskip        = 2mm,%
  keepspaces       = true,%
  mathescape       = true,%
  escapechar       = ¤,%
  tabsize          = 2,%
  numbers          = left,%
  numberstyle      = \tiny\color{Black!70},%
  numbersep        = 4pt,%
  stepnumber       = 1,%
  firstnumber      = 1,%
  showstringspaces = false,%
  captionpos       = b,%
  extendedchars    = true,%
  upquote          = true,%
  abovecaptionskip = 0mm,%
  belowcaptionskip = 0mm,%
  moredelim        = **[is][{\btHL[fill=light-gray]}]{°}{°},%
  moredelim        = **[is][{\btHL[fill=super-light-gray]}]{ç}{ç},%
  moredelim        = **[is][{\btHL[fill=lightlime]}]{§}{§},%
  moredelim        = **[is][{\btHL[fill=lightlime!70]}]{¢}{¢},%
  moredelim        = **[is][{\btHL[fill=lightpurple]}]{@}{@},%
}

\lstdefinestyle{clang}{%
  language         = C,%
  style            = basic,%
}

\newcommand{\toolname}{{\sc Trimmer}\xspace}
\newcommand{\klee}{{\sc Klee}\xspace}
\newcommand{\crab}{{\sc Crab}\xspace}
\newcommand{\kleemath}{{\sc Klee}}

\setcopyright{none}
\acmDOI{10.1145/3106237.3106249}
\acmISBN{978-1-4503-5105-8/17/09}
\acmConference[ESEC/FSE'17]{2017 11th Joint Meeting of the European Software Engineering Conference and the ACM SIGSOFT Symposium on the Foundations of Software Engineering}{September 4-8, 2017}{Paderborn, Germany}
\acmYear{2017}
\copyrightyear{2017}
\acmPrice{15.00}

\ifdefined\EXTENDEDVERSION
\else
\fi

\begin{document}

\setlength{\pdfpageheight}{\paperheight}
\setlength{\pdfpagewidth}{\paperwidth}

\ifdefined\EXTENDEDVERSION
\title{Failure-Directed Program Trimming (Extended Version)}
\else
\title{Failure-Directed Program Trimming}
\fi


\author{Kostas Ferles}
\affiliation{%
  \ifdefined\EXTENDEDVERSION
  \institution{The University of Texas at Austin}
  \city{Austin}
  \state{TX}
  \country{USA}
  \else
  \institution{The University of Texas at Austin \country{USA}}
  \fi
}
\email{kferles@cs.utexas.edu}

\author{Valentin W\"{u}stholz}
\affiliation{%
  \ifdefined\EXTENDEDVERSION
  \institution{The University of Texas at Austin}
  \city{Austin}
  \state{TX}
  \country{USA}
  \else
  \institution{The University of Texas at Austin \country{USA}}
  \fi
}
\email{valentin@cs.utexas.edu}

\author{Maria Christakis}
\affiliation{%
  \ifdefined\EXTENDEDVERSION
  \institution{University of Kent}
  \city{Canterbury}
  \country{UK}
  \else
  \institution{University of Kent \country{UK}}
  \fi
}
\email{M.Christakis@kent.ac.uk}

\author{Isil Dillig}
\affiliation{%
  \ifdefined\EXTENDEDVERSION
  \institution{The University of Texas at Austin}
  \city{Austin}
  \state{TX}
  \country{USA}
  \else
  \institution{The University of Texas at Austin \country{USA}}
  \fi
}
\email{isil@cs.utexas.edu}

\begin{abstract}
This paper describes a new program simplification technique called
\emph{program trimming} that aims to improve the scalability and precision of
safety checking tools. Given a program $\prog$, program trimming generates a
new program $\prog'$ such that $\prog$ and $\prog'$ are \emph{equi-safe}
(i.e., $\prog'$ has a bug if and only if $\prog$ has a bug), but $\prog'$ has
fewer execution paths than $\prog$. Since many program analyzers are sensitive
to the number of execution paths, program trimming has the potential to
improve the effectiveness of safety checking tools.

In addition to introducing the concept of program trimming,
this paper also presents a lightweight static analysis that can be used as a
pre-processing step to remove program paths while retaining equi-safety. We
have implemented the proposed technique in a tool called \toolname\ and
evaluate it in the context of two program analysis techniques, namely abstract
interpretation and dynamic symbolic execution. Our experiments show that
program trimming significantly improves the effectiveness
of both techniques.
\end{abstract}

\ifdefined\EXTENDEDVERSION
\else
%
%

\ccsdesc[500]{Software and its engineering~Formal methods}
\fi

\keywords{Condition inference, abstract interpretation, dynamic symbolic execution}


\maketitle

\section{Introduction}
\label{sect:intro}

Due to its potential to dramatically simplify programs with respect to a
certain criterion (e.g., the value of a program variable at a given location),
program slicing~\cite{Weiser1981} has been the focus of decades of research in
the program analysis community~\cite{Tip1995}. In addition to being useful for
program understanding, slicing also has the potential to improve the
scalability of bug-finding and verification tools by removing irrelevant code
snippets with respect to some property of interest. Yet, despite this
potential, relatively few bug-finding and verification tools use slicing as a
pre-processing step.

In this paper, we argue that existing notions of a ``program slice'' do not
adequately capture the kinds of program simplification that are beneficial to
safety checking tools. Instead, we propose a new semantic
program simplification technique called \emph{program trimming}, which removes
\emph{program paths} that are irrelevant to the safety property of
interest. Given a program $\prog$, program trimming generates a simplified
program $\prog'$ such that $\prog'$ violates a safety property if and only if
the original program $\prog$ does (i.e., $\prog$ and $\prog'$ are
\emph{equi-safe}). However, $\prog'$ has the advantage of containing fewer
execution paths than $\prog$. Since the scalability and precision of many
program analyzers depend on the number of program paths, program trimming can
have a positive impact on many kinds of program analyses, particularly those
that are \emph{not} property directed.

To illustrate the difference between the standard notion of program slicing
and our proposed notion of program trimming, consider the following very
simple program, where $\star$ indicates a non-deterministic value (e.g., user
input):
\ifdefined\EXTENDEDVERSION
\vspace{1em}
\else
\vspace{0.2em}
\fi
\begin{lstlisting}[style=clang, xleftmargin=1.2em]
x := ¤$\star$¤; y := ¤$\star$¤; ¤\label{line:init}¤
if (y > 0) { while (x < 10) { x := x + y; } } ¤\label{line:if}¤
else { x := x - 1; } ¤\label{line:else}¤
assert x > 0; ¤\label{line:assert}¤
\end{lstlisting}
\ifdefined\EXTENDEDVERSION
\vspace{0.5em}
\else
\vspace{-0.4em}
\fi

Suppose that our goal is to prove the assertion; so, we are interested in the
value of {\tt x} at line~\ref{line:assert}. Now, every single statement in
this program is relevant to determining the value of {\tt x}; hence, there is
nothing that can be removed using program slicing. However, observe that
the \code{then} branch of the \code{if} statement is actually
irrelevant to the assertion. Since this part of the program can never result
in a program state where the value of {\tt x} is less than 10,
lines~\ref{line:if} and~\ref{line:else} can be simplified without affecting
whether or not the assertion can fail. Hence, for the purposes of safety
checking, the above program is equivalent to the following much simpler
trimmed program~$\prog'$:
\ifdefined\EXTENDEDVERSION
\vspace{1em}
\else
\vspace{0.2em}
\fi
\begin{lstlisting}[style=clang, xleftmargin=1.2em]
x := ¤$\star$¤; y := ¤$\star$¤;
assume y <= 0;
x := x - 1;
assert x > 0;
\end{lstlisting}
\ifdefined\EXTENDEDVERSION
\vspace{0.5em}
\else
\vspace{-0.4em}
\fi

Observe that $\prog'$ contains far fewer paths compared to the original
program $\prog$. In fact, while $\prog$ contains infinitely many execution
paths, the trimmed program $\prog'$ contains only two, one through the
successful and one through the failing branch of the assertion. Consequently,
program analyzers that eagerly explore all program paths, such as bounded model
checkers~\cite{BiereCimatti1999,ClarkeBiere2001} and symbolic execution
engines~\cite{King1976}, can greatly benefit from program trimming in terms of
scalability. Furthermore, since many static analyzers (e.g., abstract
interpreters~\cite{CousotCousot1977}) typically lose precision at join points
of the control flow graph, program trimming can improve their precision
by removing paths that are irrelevant to a safety property.

Motivated by these observations, this paper introduces the notion of
failure-directed program trimming and presents a lightweight algorithm to
remove execution paths in a way that guarantees equi-safety. The key idea
underlying our approach is to statically infer \emph{safety conditions}, which
are sufficient conditions for correctness and can be computed in a lightweight
way. Our technique then negates these safety conditions to obtain
\emph{trimming conditions}, which are necessary conditions for the program to
fail. The trimming conditions are used to instrument the program
with assumptions such that program paths that violate an assumption are
pruned.

Program trimming is meant as a lightweight but effective pre-processing step for
program analyzers that check safety. We have implemented our proposed trimming
algorithm in a tool called \toolname\ and used it to pre-process hundreds of
programs, most of which are taken from the software verification competition
(SV-COMP)~\cite{SV-COMP}. We have also evaluated the impact of trimming in the
context of two widely-used program analysis techniques, namely abstract
interpretation~\cite{CousotCousot1977} and dynamic symbolic
execution~\cite{GodefroidKlarlund2005,CadarEngler2005}. Our experiments with
\crab~\cite{GangeNavas2016-Domain,GangeNavas2016-Sparsity} (an
abstract interpreter) show that program trimming can
considerably improve the precision of static analyzers. Furthermore, our
experiments with \klee~\cite{CadarDunbar2008} (a dynamic symbolic execution
tool) show that program trimming allows the dynamic symbolic execution engine
to find more bugs and verify more programs within a given resource limit.

To summarize, this paper makes the following key contributions:
\begin{itemize}[leftmargin=1.1em]
\item We introduce the notion of program
  trimming as a new kind of program simplification technique.
\item We propose an effective and lightweight inference engine for computing
  safety conditions.
\item We describe a modular technique for  instrumenting the program with trimming conditions.
\item We demonstrate empirically that program trimming has a significant positive impact on the effectiveness of program analyzers.
\ifdefined\EXTENDEDVERSION
  For instance, the cheapest configuration of \crab (an abstract interpreter)
with trimming
  proves 21\% more programs safe than the most expensive configuration of
  \crab without trimming in \emph{less than 70\% of the time}. In the context of a dynamic symbolic execution engine (\klee),
  trimming increases both the number of uncovered bugs by up to 30\% and the
  number of verified programs  by up to 18\% while reducing the running time
  by up to 30\%.
\fi

\end{itemize}


\section{Guided Tour}
\label{sect:tour}


The running example, shown in Figure~\ref{fig:example}, is written in C extended
with \code{assume} and \code{assert} statements. Note that the example is intentionally
quite artificial to illustrate the main ideas behind our technique. Procedure
\code{main} assigns  a non-deterministic integer value to variable \code{m} and
computes its factorial using the  recursive \code{fact} procedure. The (light and
dark) gray boxes are discussed below and \emph{should be ignored for now}. We examine
two variations of this example: one for dynamic symbolic execution (DSE)
engines and another for abstract interpreters (AI).

{\bf \emph{Motivation \#1: scalability.}}
First, let us ignore the assertion on line~\ref{line:ai-asrt} and only
consider the one on line~\ref{line:dse-asrt}. Clearly, this assertion cannot
fail unless \code{m} is equal to 123. Observe that procedure \code{main} contains
infinitely many execution paths because the number of recursive calls to
\code{fact} depends on the value of \code{m}, which is
unconstrained. Consequently, a dynamic symbolic execution engine, like
\klee, would have to explore (a number of) these paths until it finds
the bug or exceeds its resource limit. However, there is only one buggy
execution path in this program, meaning that the dynamic symbolic execution
engine is wasting its resources exploring paths that cannot possibly fail.

\begin{figure}[t!]
\begin{lstlisting}[style=clang, xleftmargin=1.2em]
int fact(int n) {
  assume 0 <= n;
  çassume n != 0;ç              // AI ¤\label{line:ai-asm}¤
  int r = 1; ¤\label{line:ai-cond}¤
  if (n != 0) {
    r = n * fact(n - 1); ¤\label{line:ai-call}¤
  }
  assert n != 0 || r == 1;    // AI ¤\label{line:ai-asrt}¤
  return r;
}

void main() {
  int m = ¤$\star$¤;
  °assume m == 123;°            // DSE ¤\label{line:dse-asm}¤
  int f = fact(m); ¤\label{line:dse-call}¤
  assert m != 123 || f == 0;  // DSE ¤\label{line:dse-asrt}¤
} ¤\label{line:dse-end}¤
\end{lstlisting}
\vspace{-1.5em}
\caption{Running example illustrating program trimming.}
\vspace{-1.5em}
\label{fig:example}
\end{figure}

{\bf \emph{Our approach.}}
Now, let us see how program trimming can help a symbolic execution tool in the
context of this example. As mentioned in Section~\ref{sect:intro}, our
program trimming technique first computes \emph{safety conditions}, which
are sufficient conditions for the rest of the program to be correct.  In this
sense, standard \emph{weakest preconditions}~\cite{Dijkstra1975} are instances
of safety conditions. However, automatically computing safety conditions
precisely, for instance via weakest precondition
calculi~\cite{Dijkstra1975,Leino2005}, abstract
interpretation~\cite{CousotCousot1977}, or predicate
abstraction~\cite{GrafSaidi1997,BallMajumdar2001}, can become very expensive
(especially in the presence of loops or recursion), making such an approach
unsuitable as a pre-processing step for program analyzers that already check
safety. Instead, we use lightweight techniques to infer safety conditions that
describe a subset of the safe executions in the program. That is, the safety
conditions inferred by our approach can be stronger than necessary, but they
are still useful for ruling out many program paths that ``obviously'' cannot
violate a safety property.

In contrast to a safety condition, a \emph{trimming condition} at a given
program point reflects a \emph{necessary} condition for the rest of the
program execution to fail. Since a necessary condition for a property $\neg Q$
can be obtained using the negation of a sufficient condition for $Q$, we can
compute a valid trimming condition for a program point $\pi$ as the negation
of the safety condition at $\pi$. Thus, our approach trims the program by
instrumenting it with assumptions of the form \code{assume} $\phi$, where
$\phi$ is the negation of the safety condition for that program point.  Since
condition $\phi$ is, by construction, necessary for the program to fail, the
trimmed program preserves the safety of the original program. Moreover,
since execution terminates as soon as we encounter an assumption violation,
instrumenting the program with trimming conditions prunes program paths in a
semantic way.

{\bf \emph{Program trimming on this example.}}
Revisiting our running example from Figure~\ref{fig:example}, the safety
condition right after line~\ref{line:dse-call} is \code{m~!=~123 || f == 0}.
Since procedure \code{fact} called at line~\ref{line:dse-call} neither
contains any assertions nor modifies the value of \code{m}, a valid safety
condition right before line~\ref{line:dse-call} is \code{m != 123}. Indeed, in
executions that reach line~\ref{line:dse-call} and satisfy this safety
condition, the assertion does not fail. We can now obtain a sound trimming
condition by negating the safety condition. This allows us to instrument the
program with the \code{assume} statement shown in the dark gray box of
line~\ref{line:dse-asm}. Any execution that does not satisfy this condition is
correct and is effectively removed by the \code{assume} statement in a way
that preserves safety. As a result, a dynamic symbolic execution tool running
on the instrumented program will only explore the single execution path
containing the bug and will not waste any resources on provably correct
paths. Observe that a bounded model checker would similarly benefit from this
kind of instrumentation.

{\bf \emph{Motivation \#2: precision.}}
To see how our approach might improve the precision of program analysis, let
us ignore the assertion on line~\ref{line:dse-asrt} and only consider the one
on line~\ref{line:ai-asrt}. Since \code{n = 0} implies \code{r = 1} on
line~\ref{line:ai-asrt}, this assertion can clearly never fail. However, an
abstract interpreter, like \crab, using intervals~\cite{CousotCousot1977}
cannot prove this assertion due to the inherent imprecision of the underlying
abstract domain. In particular, the abstract interpreter knows that \code{n}
is non-negative at the point of the assertion but has no information about
\code{r} (i.e., its abstract state is $\top$). Hence, it does not have
sufficient information to discharge the assertion at line~\ref{line:ai-asrt}.

Suppose, however, that our technique can infer the safety condition \code{n =
  0} on line~\ref{line:ai-asm}. Using this condition, we can now instrument
this line with the trimming condition \code{n != 0}, which corresponds to the
assumption in the light gray box. If we run the same abstract interpreter on
the instrumented program, it now knows that \code{n} is strictly greater than
0 and can therefore prove the assertion even though it is using the same
interval abstract domain. Hence, as this example illustrates, program trimming
can also be useful for improving the precision of static analyzers in
verification tasks.

\section{Program Trimming}
\label{sect:approach}

In this section, we formally present the key insight behind failure-directed
program trimming using a simple imperative language in the style of {\sc
  IMP}~\cite{Semantics}, augmented with \code{assert} and \code{assume}
statements. This lays the foundation for understanding the safety condition
inference, which is described in the next section and is defined for a more
expressive language. Here, we present the semantics of the IMP language using
big-step operational semantics, specifically using judgments of the form
$\pair{\sigma}{s} \Downarrow_\status \sigma'$ where:
\begin{itemize}[leftmargin=1.1em]
\item $s$ is a program statement,
\item $\sigma, \sigma'$ are \emph{valuations} mapping program variables to
  values,
\item $\varphi \in \{ \asrtfail, \asmfail, \success \}$ indicates whether an
  assertion violation occurred ($\asrtfail$), an assumption was violated
  ($\asmfail$), or neither assertion nor assumption violations were
  encountered (denoted $\success$).
\end{itemize}
We assume that the program terminates as soon as an assertion or assumption
violation is encountered.
\mv{We also ignore non-determinism  to simplify the presentation.}

\begin{definition}{\bf (Failing execution)}
We say that an execution of $s$ under $\sigma$ is \emph{failing} iff
$\pair{s}{\sigma} \Downarrow_{\asrtfail} \sigma'$, and \emph{successful}
otherwise.
\end{definition}

In other words, a failing execution exhibits an assertion
violation. Executions with \emph{assumption} violations also terminate
immediately but are not considered failing.

\begin{definition}{\bf (Equi-safety)}
We say that two programs $s, s'$ are \emph{equi-safe} iff, for all valuations
$\sigma$, we have:
\[
\pair{s}{\sigma} \Downarrow_{\asrtfail} \sigma' \ \Longleftrightarrow \ \pair{s'}{\sigma} \Downarrow_{\asrtfail} \sigma'
\]
\end{definition}

In other words, two programs are equi-safe if they exhibit the same set of
failing executions starting from the same state $\sigma$. Thus, program $s'$ has a bug
if and only if $s$ has a bug.

As mentioned in Section~\ref{sect:intro}, the goal of program trimming is to
obtain a program $s'$ that (a) is equi-safe to $s$ and (b) can terminate early
in successful executions of $s$:

\begin{definition}\label{def:trim}{\bf (Trimmed program)}
A program $s'$ is a \emph{trimmed version} of $s$ iff $s, s'$ are equi-safe and
\[
\begin{array}{ll}
(1) & \pair{s}{\sigma} \Downarrow_{\success} \sigma' \ \Longrightarrow  \pair{s'}{\sigma} \Downarrow_{\success} \sigma'
\lor \pair{s'}{\sigma} \Downarrow_{\asmfail} \sigma'' \\
(2) &  \pair{s}{\sigma} \Downarrow_{\asmfail} \sigma' \ \Longrightarrow  \pair{s'}{\sigma} \Downarrow_{\asmfail} \sigma''\\
\end{array}
\]
\end{definition}

Here, the first condition says that the trimmed program $s'$ either exhibits
the same successful execution as the original program or terminates early with
an assumption violation. The second condition says that, if the original
program terminates with an assumption violation, then the trimmed program also
violates an assumption but can terminate in a different state $\sigma''$. In
the latter case, we allow the trimmed program to end in a different state
$\sigma''$ than the original program because the assumption violation could
occur earlier in the trimmed program. Intuitively, from a program analysis
perspective, we can think of trimming as a program simplification technique
that prunes execution paths that are guaranteed not to result in an assertion
violation.

Observe that program trimming preserves all terminating executions of program
$s$. In other words, if $s$ terminates under valuation $\sigma$, then the
trimmed version $s'$ is also guaranteed to terminate. However, program
trimming does not give any guarantees about non-terminating executions. Hence,
even though this technique is suitable as a pre-processing technique for
safety checking, it does not necessarily need to preserve liveness
properties. For example, non-terminating executions of $s$ can become
terminating in $s'$.

The definition of program trimming presented above does not impose any
\emph{syntactic} restrictions on the trimmed program. For instance, it allows
program trimming to add and remove arbitrary statements as long as the
resulting program satisfies the properties of Definition~\ref{def:trim}. However, in
practice, it is desirable to make some syntactic restrictions on how trimming
can be performed. In this paper, we perform program
trimming  by adding assumptions to the original program rather
than removing statements. Even though this transformation does not ``simplify''
the program from a program understanding point of view, it is very useful to
subsequent program analyzers because the introduction of \code{assume} statements prunes
program paths in a \emph{semantic} way.

\section{Static Analysis for Trimming}
\label{sect:analysis}

As mentioned in Section~\ref{sect:intro}, our trimming algorithm consists of
two phases, where we infer safety conditions using a lightweight static
analysis in the first phase and instrument the program with trimming
conditions in the next phase. \mv{In this section, we describe the safety condition
inference.}


\subsection{Programming Language}

In order to precisely describe our trimming algorithm, we first introduce a
small, but realistic, call-by-value imperative language with pointers and
procedure calls. As shown in
Figure~\ref{fig:lang}, a program in this language consists of one or more
procedure definitions. Statements include sequencing, assignments, heap reads
and writes, memory allocation, procedure calls, assertions, assumptions, and
conditionals. Since loops can be expressed as tail-recursive procedures, we do
not introduce an additional loop construct. Also, observe that we only allow
conditionals with non-deterministic predicates, denoted $\star$. However, a
conditional of the form $\code{if } (p) \code{ \{} s_1 \code{\} else \{} s_2
\code{\}}$ can be expressed as follows in this language:
\[
\begin{array}{l}
\code{if } (\star) \code{ \{assume } p; s_1 \code{\}} \code{ else } \code{\{assume } \neg p; s_2 \code{\}}
\end{array}
\]

Since the language is quite standard, we do not present its operational
semantics in detail. However, as explained in Section~\ref{sect:approach}, we
assume that the execution of a program terminates as soon as we encounter an
assertion or assumption violation (i.e., the predicate evaluates to false). As
in Section~\ref{sect:approach}, we use the term \emph{failing execution} to
indicate a program run with an assertion violation.

\begin{figure}
\begin{grammar}
\let\syntleft\relax
\let\syntright\relax
<Program $\prog$> $::=$ $\overline{\mathit{prc}}$

<Procedure $\mathit{prc}$> $::=$ $\code{proc } \mathit{prc}(\overline{v_{\mathit{in}}}) : v_{\mathit{out}} \code { \{} s \code{\}}$

<Statement $s$> $::=$ $s_1;s_2$ | $v := e$ | $v_1 := *v_2$ | $*v := e$
 \alt $v := \code{malloc}(e)$ | $v := \code{call } \mathit{prc}(\bar{v})$
 \alt $\code{assert } p$ | $\code{assume } p$
 \alt $\code{if } (\star) \code{ \{} s_1 \code{\} else \{} s_2 \code{\}}$

<Expression $e$> $::=$ $v$ | $c$ | $e_1 \oplus e_2\ \ (\oplus \in \{+,-,\times \})$

<Predicate $p$> $::=$ $e_1 \oslash e_2\ \ (\oslash \in \{\textless,>,=\})$
  \alt $p_1 \wedge p_2$ | $p_1 \vee p_2$ | $\neg p$
\end{grammar}
\vspace{-1.2em}
\caption{Programming language used for formalization. The notation
  $\overline{s}$ denotes a sequence $s_1, \ldots, s_n$.}
\vspace{-1em}
\label{fig:lang}
\end{figure}

\subsection{Safety Condition Inference}

Recall from Section~\ref{sect:intro}, that a \emph{safety condition} at a
given program point $\pi$ is a sufficient condition for any execution starting
at $\pi$ to be error free. More precisely, a safety condition for a
(terminating) statement $s$ is a formula $\varphi$ such that $\varphi
\Rightarrow \mathit{wp}(s, {\mathit{true}})$, where $\mathit{wp}(s, \phi)$ denotes
the \emph{weakest precondition} of $s$ with respect to postcondition
$\phi$~\cite{Dijkstra1975}.  While the most precise safety condition is $\mathit{wp}(s, {\mathit{true}})$, our analysis intentionally
infers stronger safety conditions so that trimming can be used as a pre-processing technique for
safety checkers.


Our safety condition inference engine is formalized using the rules shown in
Figure~\ref{fig:rules}. Our formalization makes use of an ``oracle''
$\aliasing$ for resolving queries about pointer aliasing and procedure side
effects. For instance, this oracle can be implemented using a scalable pointer
analysis, such as the Data Structure Analysis (DSA) method of Lattner et
al.~\cite{LattnerLenharth2007}. In the rest of this section, we assume that
the oracle for resolving aliasing queries is \emph{flow-insensitive}.

Figure~\ref{fig:rules} includes two types of inference rules, one for
statements and one for procedures. Both forms of judgments utilize a
\emph{summary environment} $\summary$ that maps each procedure $\mathit{prc}$
to its corresponding safety condition (or ``summary''). Since our programming
language contains recursive procedures, we would, in general, need to perform
a fixed-point computation to obtain sound and precise summaries. However,  because our analysis initializes summaries
conservatively, the analysis can terminate at any point to produce sound
results.


With the exception of rule~(10), all  rules in Figure~\ref{fig:rules} derive judgments of the form $\aliasing, \summary, \tc \vdash
s: \tc'$. The meaning of this judgment is that, using environments $\aliasing$
and $\summary$, it is provable that $\{ \tc'\} s \{ \tc \} $ is a valid Hoare
triple (i.e., $\tc' \Rightarrow \mathit{wp}(s, \tc)$ if $s$
terminates). Similarly to the computation of standard weakest
preconditions~\cite{Dijkstra1975}, our analysis propagates safety conditions
backward but sacrifices precision to improve scalability.  In
the following discussion, we only focus on those rules where our inference
engine differs from standard precondition computation.


\begin{figure}
\[
\begin{array}{cc}
(1) & \irule{
\begin{array}{c}
\aliasing, \summary, \tc \vdash s_2: \tc_2 \\
\aliasing, \summary, \tc_2 \vdash s_1: \tc_1
\end{array}
}{\aliasing, \summary, \tc \vdash s_1; s_2: \tc_1} \\ \ \\
(2) & \irule{
\tc' \equiv \tc[e/v]
}
{\aliasing, \summary, \tc \vdash v := e: \tc'}  \\ \ \\
(3) & \irule{
\tc' \equiv \tc[\mathit{drf}(v_2)/v_1]
}
{\aliasing, \summary, \tc \vdash v_1 := *v_2: \tc'}  \\ \ \\
(4) & \irule{
\tc' \equiv \mathit{store}(\mathit{drf}(v), e, \aliasing, \tc)
}
{\aliasing, \summary, \tc \vdash *v := e: \tc'}  \\ \ \\
(5) & \irule{
\tc' \equiv \forall v.  \tc
}
{\aliasing, \summary, \tc \vdash v := \code{malloc}(e): \tc'}  \\ \ \\
(6) & \irule{
\begin{array}{c}
\overline{\alpha} \equiv \mathit{modLocs}(\mathit{prc}, \aliasing) \\
\tc_s \equiv \forall v. \ \mathit{havoc}(\overline{\alpha}, \aliasing,  \tc) \\
\tc' \equiv \tc_s \wedge \mathit{summary}(\mathit{prc}, \summary, \overline{v_\mathit{act}})
\end{array}
}
{\aliasing, \summary, \tc \vdash {v} := \code{call } \mathit{prc}(\overline{v_\mathit{act}}): \tc'}  \\ \ \\
(7) & \irule{
\tc' \equiv p \wedge \tc
}
{\aliasing, \summary, \tc \vdash \code{assert } p: \tc'}  \\ \ \\
(8) & \irule{
\tc' \equiv p \Rightarrow \tc
}
{\aliasing, \summary, \tc \vdash \code{assume } p: \tc'}  \\ \ \\
(9) & \irule{
\begin{array}{c}
\aliasing, \summary, \tc \vdash s_1: \tc_1 \\
\aliasing, \summary, \tc \vdash s_2: \tc_2 \\
\tc' \equiv \tc_1 \wedge \tc_2
\end{array}
}{\aliasing, \summary, \tc \vdash \code{if } (\star) \code{ \{} s_1 \code{\} else \{} s_2 \code{\}}: \tc' } \\ \ \\
(10) & \irule{
\begin{array}{c}
\aliasing, \summary, \mathit{true} \vdash s: \tc \\
\summary' \equiv \summary[\mathit{prc} \mapsto \tc]
\end{array}
}{\aliasing, \summary \vdash \code{proc } \mathit{prc}(\overline{v_{\mathit{in}}}) : v_{\mathit{out}} \code { \{} s \code{\}}: \summary' }
\end{array}
\]
\vspace{-1.3em}
\caption{Inference rules for computing safety conditions.}
\vspace{-1.2em}
\label{fig:rules}
\end{figure}

{\bf \emph{Heap reads and writes.}}
\mv{An innovation underlying our safety condition inference is the
handling of the heap.  Given a store operation $*v := e$, this
statement can modify the value of all expressions $*x$, where $x$ is an alias
of $v$. Hence, a sound way to model the heap is to rewrite $*v := e$ as}
\[
*v := e; \code{if }(v = v_1) \ *v_1 := e; \ldots; \code{if }(v = v_k) \ *v_k := e;
\]
where $v_1, \ldots, v_k$ are potential aliases of $v$. Effectively, this
strategy
accounts for the ``side effects'' of statement $*v := e$ to other heap
locations by explicitly introducing additional statements. These
statements are of the form $\code{if} \ (v = v_i)$ $*v_i := e$, i.e., if $v$ and
$v_i$ are indeed aliases, then change the value of expression $*v_i$ to $e$.

While the strategy outlined above is sound, it unfortunately conflicts with
our goal of computing safety conditions using lightweight analysis. In
particular, since we use a coarse, but scalable alias analysis, most pointers
have a large number of possible aliases in practice. Hence, introducing a
linear number of conditionals causes a huge blow-up in the size of the safety
conditions computed by our technique. To prevent this blow-up, our inference
engine computes a safety precondition that is stronger than necessary by using
the following conservative $\mathit{store}$ operation.

\begin{definition}{{\bf (Memory location)}}
We represent memory locations using terms that belong to the following
grammar:
\[
\begin{array}{ccc}
\emph{Memory location} \ \alpha := v \ | \ \mathit{drf}(\alpha)
\end{array}
\]
Here, $v$ represents any program variable, and $\mathit{drf}$ is an
uninterpreted function representing the dereference of a memory location.
\end{definition}

To define our conservative store operation, we make use of a function
$\mathit{aliases}(v, \aliasing)$ that uses oracle $\aliasing$ to retrieve all
memory locations $\alpha$ that may alias $v$.

\begin{definition}{\bf (Store operation)}
Let $\mathit{derefs}(\tc)$ denote all $\alpha'$ for which a sub-term
$\mathit{drf}(\alpha')$ occurs in formula $\tc$. Then,
\begin{align*}
\mathit{store}(\mathit{drf}(\alpha), e, \aliasing, \tc) := \tc[e/\mathit{drf}(\alpha)] \wedge
\bigwedge_{\alpha_i \in A \setminus \{ \alpha \}} \alpha_i \neq \alpha \\
\textrm{where} \ A \equiv \mathit{aliases}(\alpha, \aliasing)  \cap \mathit{derefs}(\tc)
\end{align*}
\end{definition}

In other words, we compute the precondition for statement $*v := e$ as though
the store operation was a regular assignment, but we also ``assert'' that $v$
is distinct from every memory location $\alpha_i$ that can potentially alias
$v$. To see why this is correct, observe that $\tc[e/\mathit{drf}(v)]$ gives the
weakest precondition of $*v := e$ when $v$ does not have any aliases. If $v$
does have aliases that are relevant to the safety condition, then the conjunct
$\bigwedge_{\alpha_i \in A \setminus \{ v \}} \alpha_i \neq v$ evaluates to
$\mathit{false}$, meaning that we can never guarantee the safety of the program. Thus,
$\mathit{store}(\mathit{drf}(v), e, \aliasing, \tc)$ logically implies
$\mathit{wp}(*v := e, \tc)$.


\begin{example}
Consider the following code snippet:
\textnormal{
\[
\begin{array}{l}
\code{if }(\star) \ \code{\{} \code{assume } x = y;  a := 3; \code{\}} \\
\code{else } \ \ \code{\{} \code{assume } x \neq y; *y := 3; \code{\}} \\
*x := a; t:= *y; \\
\code{assert } t = 3;
\end{array}
\]
}
Right before the heap write $*x := a$, our analysis infers the safety
condition $\mathit{drf}(y) = 3 \land x \neq y$. Before the heap write $*y :=
3$, the safety condition is ${x \neq y}$, which causes the condition before the
assumption $\code{assume } x \neq y$  to be $\mathit{true}$. This means that
executions through the \code{else} branch are verified and may be trimmed
because $x$ and $y$ are not aliases for these executions.
\end{example}

{\bf \emph{Interprocedural analysis.}}
We now turn our attention to the handling of procedure calls. As mentioned
earlier, we perform interprocedural analysis in a modular way, computing
summaries for each procedure. Specifically, a summary $\summary(f)$ for
procedure $f$ is a sufficient condition for any execution of $f$ to be error
free.

With this intuition in mind, let us consider rule~(6) for analyzing procedure
calls of the form ${v} := \code{call } \mathit{prc}(\bar{e})$. Suppose that
$\bar{\alpha}$ is the set of memory locations modified by the callee
$\mathit{prc}$ but expressed in terms of the memory locations in the
\emph{caller}. Then, similarly to other modular interprocedural
analyses~\cite{BarnettChang2005,BarnettFahndrich2011}, we conservatively model
the effect of the statement ${v} := \code{call }
\mathit{prc}(\overline{v_\mathit{act}})$ as follows:
\[
\begin{array}{l}
\code{assert }\mathit{summary}(\mathit{prc}); \\
\code{havoc } v; \code{havoc }\bar{\alpha};
\end{array}
\]
Here, \code{havoc} $\alpha$ denotes a statement that assigns an unknown value
to memory location $\alpha$. Hence, our treatment of procedure calls asserts
that the safety condition for $\mathit{prc}$ holds before the call and that
the values of all memory locations modified in $\mathit{prc}$ are
``destroyed''.

While our general approach is similar to prior techniques on modular
analysis~\cite{BarnettChang2005,BarnettFahndrich2011}, there are some
subtleties in our context to which we would like to draw the reader's
attention. First, since our procedure summaries (i.e., safety conditions) are
not provided by the user, but instead inferred by our algorithm (see rule~(10)), we must be
conservative about how summaries are ``initialized''. In particular, because
our analysis aims to be lightweight, we do not want to perform an expensive
fixed-point computation in the presence of recursive procedures. Therefore, we use the following $\mathit{summary}$ function to yield a
conservative summary for each procedure.

\begin{definition} {\bf (Procedure summary)}
Let $\mathit{hasAsrts}(f)$ be a predicate that yields $\mathit{true}$ iff procedure $f$
or any of its (transitive) callees contain an assertion. Then,
\[
\mathit{summary}(\mathit{f}, \summary, \bar{v}) =
\left \{
\begin{array} {ll}
\summary(f)[\bar{v}/\overline{v_\mathit{in}}] & \emph{if} \ f \in \mathit{dom}(\summary) \\
\mathit{false} & \emph{if} \ \mathit{hasAsrts}(f) \\
\mathit{true} & \emph{otherwise}
\end{array}
\right .
\]
\end{definition}

In other words, if procedure $f$ is in the domain of $\summary$ (meaning that
it has previously been analyzed), we use the safety condition given by
$\summary(f)$, substituting formals by the actuals. However, if $f$ has not
yet been analyzed, we then use the conservative summary $\mathit{false}$ if
$f$ or any of its callees have assertions, and $\mathit{true}$
otherwise. Observe that, if $f$ is not part of a strongly connected component
(SCC) in the call graph, we can always obtain the precise summary for $f$ by
analyzing the program bottom-up. However, if $f$ is part of an SCC, we can
still soundly analyze the caller by using the conservative summaries given by
$\mathit{summary}(\mathit{f}, \summary, \bar{v})$.

The other subtlety about our interprocedural analysis is the particular way in
which havocking is performed. Since the callee may modify heap locations
accessible in the caller, we define a $\mathit{havoc}$ operation that uses the
$\mathit{store}$ function from earlier to conservatively deal with memory locations.

\begin{definition}{\bf (Havoc operation)}
\begin{align*}
\mathit{havoc}(\mathit{drf}(\alpha), \aliasing, \tc) := \ &\forall v_\mathit{new} . \ \mathit{store}(\mathit{drf}(\alpha), v_\mathit{new}, \aliasing, \tc)\\
&\textrm{where} \ v_\mathit{new} \notin \mathit{freeVars}(\tc)\\
\mathit{havoc}(\overline{\alpha}, \aliasing, \tc) := \ &\mathit{havoc}(\mathit{tail}(\overline{\alpha}), \aliasing, \mathit{havoc}(\mathit{head}(\overline{\alpha}), \aliasing, \tc))\\
\end{align*}
\vspace{-3em}
\end{definition}

Observe that the above definition differs from the standard way this operation
is typically defined~\cite{BarnettChang2005}. In particular, given a scalar
variable $v$, the assignment $v := \star$, and its postcondition $\phi$, the
standard way to compute a conservative precondition for the assignment is
$\forall v. \phi$ (i.e., $\phi$ must hold for \emph{any} value of $v$). Note
that an alternative way of computing the precondition is $\forall
x. \phi[x/v]$, where $x$ is not a free variable in $\phi$. In the context of
scalars, these two definitions are essentially identical, but the latter view
allows us to naturally extend our definition to heap locations by using the
previously defined $\mathit{store}$ function. Specifically, given a heap
location $\mathit{drf}(\alpha)$ modified by the callee, we model the effect of
this modification as $\forall v_\mathit{new}
. \ \mathit{store}(\mathit{drf}(\alpha), v_\mathit{new}, \aliasing, \tc)$.

\begin{theorem}
\label{THM:TC}
Suppose that $\aliasing, \summary, \tc \vdash s: \tc'$, and assume that
$\aliasing$ provides sound information about aliasing and procedure side
effects. Then, under the condition that $s$ terminates and that the summaries
provided by $\summary$ are sound, we have $\tc' \Rightarrow \mathit{wp}(s,
\tc)$.
\ifdefined\EXTENDEDVERSION
\footnote{Proofs or proof sketches for all theorems can be found in the appendix.}
\else
\fi
\end{theorem}

\section{Program Instrumentation}
\label{sect:instrumentation}

In the previous section, we discussed how to infer safety conditions for each
program point. Recall that program trimming annotates the code
with \emph{trimming conditions}, which are necessary conditions for
failure. Here, we describe how we instrument the program with
suitable assumptions that preserve safety of the original program.

{\bf \emph{Intraprocedural instrumentation.}}
First, let us ignore procedure calls and consider instrumenting a single
procedure in isolation. Specifically, consider a procedure with body $s_1;
\ldots; s_n$ and let:
\vspace{-0.1in}
\[
\aliasing, \summary, \mathit{true} \vdash s_{i}; \ldots; s_n: \tc
\]
We instrument the program with the statement $\code{assume }\neg \tc$ right
before statement $s_i$ if $s_i$ complies with the instrumentation strategy
specified by the user (see Section~\ref{sect:implementation}).
In general, note that we do not instrument at every
single instruction because subsequent safety checkers must also analyze the
assumptions, which adds overhead to their analysis.

\begin{theorem}
\label{THM:INSTR}
Suppose that our technique adds a statement \textnormal{\code{assume}} $\tc$
before $s_i; \ldots; s_n$. Then, $\tc$ is a necessary condition for $s_i;
\ldots; s_n$ to have an assertion violation.
\end{theorem}

{\bf \emph{Interprocedural instrumentation.}}
One of the key challenges in performing program instrumentation is how to
handle procedure calls. In particular, we cannot simply annotate a procedure
$f$ using the safety conditions computed for $f$. The following example
illustrates why such a strategy would be unsound.

\vspace{1em}
\begin{example}~\label{ex:instr}
Consider procedures \textnormal{\code{foo}},
\textnormal{\code{bar}}, and \textnormal{\code{baz}}:
\textnormal{
\[
\vspace{1.5em}
\begin{array}{l}
\code{proc foo}(x) \ \code{\{} \mathit{*x} := 2; \code{\}} \\
\code{proc bar}(a) \ \code{\{} x := \code{malloc}(a); \code{foo}(x); \code{assert } a < 100; \code{\}} \\
\code{proc baz}(b) \ \code{\{} x := \code{malloc}(b); \code{foo}(x); \code{assert } b > 10;  \code{\}}
\end{array}
\vspace{-1em}
\]
}
Here, the safety condition for procedure \textnormal{\code{foo}} is just
$\mathit{true}$ since \textnormal{\code{foo}} does not contain assertions or
have callees with assertions. However, observe that we cannot simply
instrument \textnormal{\code{foo}} with \textnormal{\code{assume}}
$\mathit{false}$ because there are assertions after the call to
\textnormal{\code{foo}} in \textnormal{\code{bar}} and
\textnormal{\code{baz}}.
One possible solution to this challenge is to only instrument
the \textnormal{\code{main}} method, which would be very ineffective. Another possible strategy might be to
propagate safety conditions top-down from callers to callees in a separate
pass. However, this latter  strategy also has some drawbacks. For instance, in this example,
variables
\textnormal{\code{a}} and \textnormal{\code{b}} are not in scope in
\textnormal{\code{foo}}; hence, there is no meaningful instrumentation we
could add to \textnormal{\code{foo}} short of \textnormal{\code{assume}}
$\mathit{true}$, which is the same as having no instrumentation at all.
\end{example}

We solve this challenge by performing a program transformation inspired by
previous work~\cite{GurfinkelWei2008,LalQadeer2014-MixedSemantics}. The key
idea underlying this program transformation is to create, for each procedure
$\mathit{prc}$, a new procedure $\mathit{prc'}$ that can never fail. In
particular, we create $\mathit{prc'}$ by (a)~changing all assertions
\code{assert} $\phi$ in $\mathit{prc}$ to \code{assume} $\phi$, and
(b)~replacing all calls to $\mathit{f}$ (including recursive ones) with
$\mathit{f'}$. Now, given a call site of $\mathit{prc}$, $v := \code{call }
\mathit{prc}(\bar{e})$, we replace it with the following conditional:
\textnormal{
\[
\begin{array}{l}
\code{if } (\star) \code{ \{} v := \code{call } \mathit{prc}'(\bar{e}); \code{\}} \\
\code{else } \ \ \ \code{\{} v := \code{call } \mathit{prc}(\bar{e}); \code{assume} \  \mathit{false}; \code{\}}
\end{array}
\vspace{-0.5em}
\]
}

This transformation is semantics preserving since it is merely a case
analysis: Either $\mathit{prc}$ succeeds, in which case it is safe to replace
the call to $\mathit{prc}$ with $\mathit{prc'}$, or it fails, in which case we
can call original $\mathit{prc}$ but add \code{assume} $\mathit{false}$
afterward since $\mathit{prc}$ has failed. The following example illustrates
this transformation.

\begin{example}
Consider the following procedures:
%
\textnormal{
\vspace{-0.3em}
\[
\begin{array}{l}
\code{proc foo}(x, y) \ \ \ \code{\{} \code{assert } x > 0; \code{bar}(y); \code{\}} \\
\code{proc bar}(z) \ \ \ \ \ \ \ \code{\{} \code{assert } z > 0; \code{\}} \\
\code{proc main}(x, y) \ \code{\{} \code{foo}(x, y);  \code{\}}
\end{array}
\]
}
Our transformation yields the following new program:
\textnormal{
\[
\begin{array}{l}
\code{proc foo'}(x, y) \ \ \code{\{} \code{assume } x > 0; \code{bar'}(y); \code{\}}  \\
\code{proc foo}(x, y) \ \ \ \ \code{\{} \\
\ \ \ \ \code{assert } x > 0;  \\
\ \ \ \ \code{if }(\star) \ \code{\{} \code{bar'}(y); \code{\}}\\
\ \ \ \ \code{else} \ \ \ \ \code{\{} \code{bar}(y); \code{assume } \mathit{false}; \code{\}}  \\
\code{\}}  \\
\code{proc bar'}(z) \ \ \ \ \ \ \code{\{} \code{assume } z > 0; \code{\}} \\
\code{proc bar}(z) \ \ \ \ \ \ \ \ \code{\{} \code{assert } z > 0; \code{\}} \\
\code{proc main}(x, y) \ \ \code{\{} \\
\ \ \ \ \code{if }(\star) \ \code{\{} \code{foo'}(x, y); \code{\}}\\
\ \ \ \ \code{else} \ \ \ \ \code{\{} \code{foo}(x, y); \code{assume } \mathit{false}; \code{\}} \\
\code{\}}
\end{array}
\]
}
\end{example}

The main advantage of this transformation is that it allows us to perform
program instrumentation in a modular and conceptually simple way. In
particular, we do not need to instrument the ``safe'' version $\mathit{prc'}$
of a procedure $\mathit{prc}$ since $\mathit{prc'}$ never fails. On the other
hand, it is safe to instrument $\mathit{prc}$ with the negation of the local
safety conditions since every call site of $\mathit{prc}$ is followed by the
statement \code{assume} $\mathit{false}$ (i.e., execution terminates
immediately after the call).

\begin{example}
Consider the following procedures \textnormal{\code{foo}} and
\textnormal{\code{bar}}:
\textnormal{
\[
\begin{array}{l}
\code{proc foo}(x) \ \ \ \ \, \code{\{} \code{assert } x > 10; \code{\}} \\
\code{proc bar}(a, x) \ \code{\{} \code{foo}(x); \code{assert } a < 100; \code{\}} \\
\end{array}
\]
}
Our instrumentation yields the following new program:
\textnormal{
\[
\begin{array}{l}
\code{proc foo'}(x) \ \ \, \code{\{} \code{assume } x > 10; \code{\}} \\
\code{proc foo}(x) \ \ \ \ \, \code{\{} \code{assume } x \leq 10; \code{assert } x > 10; \code{\}} \\
\code{proc bar}(a, x) \ \code{\{} \\
\ \ \ \ \code{assume } a \geq 100 \lor x \leq 10; \\
\ \ \ \ \code{if }(\star) \ \code{\{} \code{foo'}(x); \code{\}} \\
\ \ \ \ \code{else} \ \ \ \ \code{\{} \code{foo}(x); \code{assume } \mathit{false}; \code{\}} \\
\ \ \ \ \code{assert } a < 100; \\
\code{\}} \\
\end{array}
\]
}
\end{example}

{\bf \emph{Discussion.}}
The reader may notice that our program transformation introduces additional
branches that did not exist in the original program. Since the goal of program
trimming is to reduce the number of execution paths while retaining
equi-safety, this transformation may seem counter-intuitive.  However, because
one of the branches is always followed by \code{assume} $\mathit{false}$, our
transformation does not lead to a blow-up in the number of paths and allows us
to perform the instrumentation modularly.

\section{Implementation}
\label{sect:implementation}

We have implemented our program trimming algorithm as a tool called \toolname,
meant as a lightweight pre-processor for program analyzers that check
safety. Our implementation is
based on the LLVM infrastructure~\cite{LattnerAdve2004} and performs
instrumentation at the LLVM bit-code level. Hence, \toolname can be
conveniently integrated into any safety checking tool that is built on top of
the LLVM infrastructure and is capable of analyzing \code{assume} statements.

Recall from Section~\ref{sect:analysis} that \toolname's safety inference
engine requires alias and side effect information to soundly analyze heap
stores and procedure calls. For this purpose, \toolname leverages LLVM's DSA
pointer analysis~\cite{LattnerLenharth2007}, a highly-scalable,
summary-based, flow-insensitive analysis.

Since \toolname can be useful to a variety of program analysis tools
(including both static and dynamic analyzers), \toolname can be customized in
different ways depending on the assumptions made by subsequent safety
checkers. In what follows, we describe the different configurations that
\toolname provides.

\emph{Reasoning about integer arithmetic.}
\toolname provides the option of treating integral-type expressions either as
mathematical (unbounded) or fixed-width integers.  Since some safety checkers
ignore integer over- and under-flows but others do not, \toolname supports
both encodings.\footnote{For the fixed-width integer encoding, \toolname
  strengthens safety conditions by requiring that there are no integer over-
  or under-flows. Specifically, \toolname utilizes arithmetic operations in
  the LLVM instruction set that return both the result of the operation and a
  flag indicating whether an over-flow occurred. Note that \toolname does not
  use bit-vectors for encoding fixed-width integers.} Analyzers treating
values as mathematical integers can therefore use the configuration of
\toolname that also makes this same unsound assumption.

\emph{Eliminating quantifiers.}
Recall from Section~\ref{sect:analysis} that the safety conditions generated
by our inference engine contain universal quantifiers. Hence, when negating
the safety conditions, the resulting trimming conditions contain
existentially-quantified variables. \toolname provides two alternatives for
eliminating quantifiers. First, \toolname can remove quantifiers using Z3's
quantifier elimination (QE) capabilities~\cite{deMouraBjorner2008} after
simplifying and pre-processing the formula. Second, \toolname also allows
replacing quantified variables by calls to non-deterministic functions. Since
quantified variables at the formula level correspond to program variables with
unknown values, this strategy has the same effect as quantifier elimination.

\emph{Bounding the instrumentation.}
After \toolname instruments the program with trimming conditions, subsequent
safety checkers need to analyze the assumptions. Hence, the number of
additional \code{assume} statements as well as the \emph{size} of the
predicates can affect the running time of program analyzers. For this reason,
\toolname allows users to customize where to add assumptions in the code. For
example, sensible strategies include adding instrumentation right before loops
and procedure calls, or before every conditional.

In a similar vein, \toolname also provides different options for bounding the
size of the formulas used in \code{assume} statements. For example, the user
can bound the number of conjuncts in the formula to be at most $k$, where $k$
is a value chosen by the user. This strategy is sound because \toolname
guarantees that the ``simplified'' formulas are weaker than the original
trimming conditions.

\section{Experiments}
\label{sect:experiments}

To evaluate the effectiveness of program trimming, we have
used \toolname to pre-process hundreds of programs by instrumenting them with
\code{assume} statements. Since these assumptions are not useful on their own,
we evaluate the effect of program trimming in the context of two different
LLVM-based program analyzers for safety checking. In particular, we use \crab,
an abstract interpreter that supports several abstract domains, and \klee, a
widely-used dynamic symbolic execution engine.

We ran our experiments on 439 programs\footnote{Available
    at: \url{https://mariachris.github.io/FSE2017/benchmarks.zip}}, most of which (92\%) are taken from
the software verification competition (SV-COMP) benchmarks, \mv{which have
  clearly defined outcomes and are handled by numerous tools}.  Since the
errors in many of the buggy programs in this benchmark set are very
shallow\footnote{For example, in the existing SV-COMP benchmarks, \klee can
  find the bug with a very low resource limit for 85\% of the buggy
  programs.}, we also augment these benchmarks with additional buggy programs,
either taken from other sources or obtained by injecting deeper bugs into
\mv{safe} SV-COMP benchmarks. The benchmarks taken from SV-COMP span a broad range of
categories, including \textsc{ControlFlow}, \textsc{Loops},
\textsc{Recursive}, and \textsc{ArrayReach}, \mv{but exclude categories that are
  not handled by  \klee or \crab, e.g., \textsc{BitVectorsReach},
  \textsc{Concurrency}}. 

In what follows, we describe the effects of program trimming on the results of
\crab and \klee. We ran all of our experiments on an Intel Xeon CPU E5-2640 v3
@ 2.60GHz machine with 132 GB of memory running the Ubuntu 14.04.1 operating
system. \mv{We used the latest available version of \crab and the latest version of \klee that was
  compatible with LLVM 3.6, which \crab requires.}

\subsection{Impact of Program Trimming on \crab}

To demonstrate that program trimming increases precision across a range of
abstract domains, we compare the performance of \crab (with and without
trimming) on three different domains with varying levels of precision:

\begin{itemize}[leftmargin=1.1em]
\item {\bf Int} denotes the (non-relational) interval
  domain~\cite{CousotCousot1977}, which infers invariants of the form $c_1
  \leq x \leq c_2$;
\item {\bf Zones} is the (relational) zones abstract domain~\cite{Mine2004},
  which infers difference constraints of the form $x-y \leq c$;
\item {\bf RTZ} is \crab's most precise (native) abstract domain and
  corresponds to the reduced product of disjunctive intervals (i.e.,
  disjunctions of constraints of the form $c_1 \leq x \leq
  c_2$)~\cite{FahndrichLogozzo2010} and the zones abstract domains.
\end{itemize}

As mentioned in Section~\ref{sect:implementation}, \toolname can be customized
using a variety of different configurations. To understand the precision
vs. performance trade-off, we evaluate \crab using the configurations of
\toolname shown in Table~\ref{tab:configs}.  Here, the column labeled {\sc MC}
indicates the maximum number of conjuncts used in an \code{assume}
statement. The third column labeled {\sc QE} indicates whether we use
quantifier elimination or whether we model quantified variables using calls to
non-deterministic functions (recall
Section~\ref{sect:implementation}). Finally, the columns labeled {\sc L/P} and
{\sc C} denote the instrumentation strategy. In configurations where there is
a checkmark under {\sc L/P}, we add \code{assume} statements right before
loops (L) and before procedure (P) calls. In configurations where there is a
checkmark under C, we also add instrumentation before every conditional.  The
two \mv{right-most} columns show the total number of added \code{assume} statements
(not trivially $\mathit{true}$) and the pre-processing time for all benchmarks. Since average
trimming time is 11--20 milliseconds per benchmark, we see that program trimming is indeed very lightweight.


\begin{table}[t!]
\centering
\caption{Overview of trimming configurations (incl. total number of added \code{assume} statements and time for pre-processing all benchmarks in the two \mv{right-most} columns). }
\vspace{-0.7em}
\scalebox{0.95}{
\begin{tabular}{l|c|c|c|c||r|r}
\textsc{Configuration} & \textsc{MC} & \textsc{QE} & \textsc{L/P} & \textsc{C} & \textsc{A} & \textsc{Time (s)}\\
\hline
\textbf{Trim$_{L+B}$}  & 4 & \yes & \yes & \no  & 143 & 5.31\\
\textbf{Trim$_{B}$}    & 4 & \yes & \yes & \yes & 1638 & 4.97\\
\textbf{Trim$_{\mathit{ND}+B}$} & 4 & \no & \yes & \yes & 2801 & 7.34\\
\textbf{Trim$_{L}$}    & $\infty$ & \yes & \yes & \no & 156 & 6.05\\
\textbf{Trim}         & $\infty$ & \yes & \yes & \yes & 1735 & 5.74\\
\textbf{Trim$_{\mathit{ND}}$}   & $\infty$ & \no & \yes & \yes & 2852 & 8.62\\
\end{tabular}
}
\vspace{-1em}
\label{tab:configs}
\end{table}

The results of our evaluation are summarized in Table~\ref{tab:crab}. As we
can see from this table, all configurations of program trimming improve the
precision of \crab, and these improvements range from 23\% to 54\%. For
instance, for the interval domain, the most precise configuration of \toolname
allows the verification of 68 benchmarks instead of only 49 when using \crab
without trimming.

Another observation based on Table~\ref{tab:crab} is the precision
vs. performance trade-offs between different configurations of
\toolname. Versions of \crab that use \toolname with QE seem to be
faster and more precise than those configurations of \toolname without QE.
In particular, the version of \toolname with QE performs better because
there are fewer variables for the abstract domain to track. We also conjecture
that \toolname using QE is more precise because the abstract domain can
introduce imprecision when reasoning about logical connectives. For instance,
consider the formula $\exists x. (x = 1 \land x \neq 1)$, which is logically
equivalent to $\mathit{false}$, so \toolname with QE would
instrument the code with \code{assume} $\mathit{false}$. However, if we do not
use QE, we would instrument the code as follows:
\[
x := \code{nondet}(); \code{assume } x = 1 \land x \neq 1;
\]
When reasoning about the \code{assume} statement, an abstract interpreter
using the interval domain takes the meet of the intervals $[1, 1]$ and $\top$,
which yields $[1, 1]$. Hence, using \toolname without QE, \crab cannot prove
that the subsequent code is unreachable.

\begin{table}[t!]
\centering
\caption{Increased precision of an abstract interpreter due to trimming. Since
  \crab treats integers as unbounded, our instrumentation also makes this
  assumption.}
\vspace{-0.7em}
\scalebox{0.95}{
\begin{tabular}{l|r|r}
\textsc{Configuration} & \textsc{Safe} & \textsc{Time (s)}\\
\hline
\textbf{Int} & 49 (+0\%) & 129 (+0\%)\\
\textbf{Trim$_{L+B}$ + Int} & 63 (+29\%) & 149 (+16\%)\\
\textbf{Trim$_{B}$ + Int} & 65 (+33\%) & 173 (+34\%)\\
\textbf{Trim$_{\mathit{ND}+B}$ + Int} & 61 (+24\%) & 198 (+53\%)\\
\textbf{Trim$_{L}$ + Int} & 64 (+31\%) & 151 (+17\%)\\
\textbf{Trim + Int} & \textbf{68 (+39\%)} & 191 (+48\%)\\
\textbf{Trim$_{\mathit{ND}}$ + Int} & 62 (+27\%) & 227 (+76\%)\\
\hline
\textbf{Zones} & 52 (+0\%) & 130 (+0\%)\\
\textbf{Trim$_{L+B}$ + Zones} & 66 (+27\%) & 148 (+14\%)\\
\textbf{Trim$_{B}$ + Zones} & 68 (+31\%) & 195 (+50\%)\\
\textbf{Trim$_{\mathit{ND}+B}$ + Zones} & 64 (+23\%) & 222 (+71\%)\\
\textbf{Trim$_{L}$ + Zones} & 67 (+29\%) & 150 (+15\%)\\
\textbf{Trim + Zones} & \textbf{73 (+40\%)} & 281 (+116\%)\\
\textbf{Trim$_{\mathit{ND}}$ + Zones} & 66 (+27\%) & 320 (+146\%)\\
\hline
\textbf{RTZ} & 52 (+0\%) & 215 (+0\%)\\
\textbf{Trim$_{L+B}$ + RTZ} & 67 (+29\%) & 231 (+7\%)\\
\textbf{Trim$_{B}$ + RTZ} & 76 (+46\%) & 535 (+149\%)\\
\textbf{Trim$_{\mathit{ND}+B}$ + RTZ} & 66 (+27\%) & 582 (+171\%)\\
\textbf{Trim$_{L}$ + RTZ} & 68 (+31\%) & 237 (+10\%)\\
\textbf{Trim + RTZ} & \textbf{80 (+54\%)} & 1620 (+653\%)\\
\textbf{Trim$_{\mathit{ND}}$ + RTZ} & 67 (+29\%) & 3330 (+1449\%)\\
\end{tabular}
}
\vspace{-1.3em}
\label{tab:crab}
\end{table}

\emph{Summary.}
Table~\ref{tab:crab} shows that trimming
significantly improves the precision of an abstract interpreter with
reasonable overhead. Our cheapest trimming configuration
(\textbf{Trim$_{L+B}$ + Int}) proves 21\% more programs safe than the most
expensive configuration of \crab without trimming (\textbf{RTZ}) in \emph{less
  than 70\% of the time}.

\subsection{Impact of Program Trimming on \klee}

In our second experiment, we evaluate the impact of program trimming on \klee,
a state-of-the-art dynamic symbolic execution tool. We use a subset\footnote{In particular, since \klee's analysis is already path-sensitive we do not consider variants that instrument before conditionals here.} of the
variants of \toolname (see Table~\ref{tab:configs}) and
evaluate trimming on \klee with three search strategies:
breadth-first search~(BFS), depth-first search~(DFS), and random search~(R).

\begin{table*}[t!]
\centering
\caption{Summary of comparison with \klee. Since \klee treats integers in a
  sound way, we also use the variant of \toolname that reasons about integer
  over- and under-flows.}
\vspace{-0.5em}
\scalebox{0.95}{
\begin{tabular}{l|r|r|r|r|r|c|r}
\textsc{Configuration} & \textsc{Safe} & \textsc{Unsafe} & \textsc{Paths} & \textsc{Timeout} & \textsc{Max-Forks} & \textsc{Time (s)}\\
\hline
\textbf{\kleemath$_{\mathit{BFS}}$} & 126 (+0\%) & 118 (+0\%) & 9231 (+0\%) & 73 (+0\%) & 73 (+0\%) & 21679\\
\textbf{Trim$_{L+B}$ + \kleemath$_{\mathit{BFS}}$} & 146 (+16\%) & 145 (+23\%) & 5978 (-35\%) & 52 (-40\%) & 46 (-51\%) & 15558\\
\textbf{Trim$_{L}$ + \kleemath$_{\mathit{BFS}}$} & \textbf{146 (+16\%)} & \textbf{153 (+30\%)} & \textbf{5678 (-38\%)} & \textbf{50 (-32\%)} & \textbf{40 (-45\%)} & \textbf{15264}\\
\hline
\textbf{\kleemath$_{\mathit{DFS}}$} & 126 (+0\%) & 99 (+0\%) & 10024 (+0\%) & 91 (+0\%) & 75 (+0\%) & 26185\\
\textbf{Trim$_{L+B}$ + \kleemath$_{\mathit{DFS}}$} & 146 (+16\%) & 124 (+25\%) & 6939 (-31\%) & 72 (-21\%) & 48 (-36\%) & 20797\\
\textbf{Trim$_{L}$ + \kleemath$_{\mathit{DFS}}$} & \textbf{146 (+16\%)} & \textbf{129 (+30\%)} & \textbf{6695 (-33\%)} & \textbf{72 (-21\%)} & \textbf{43 (-43\%)} & \textbf{21164}\\
\hline
\textbf{\kleemath$_{R}$} & 126 (+0\%) & 121 (+0\%) & 9227 (+0\%) & 71 (+0\%) & 72 (+0\%) & 21077\\
\textbf{Trim$_{L+B}$ + \kleemath$_{R}$} & 149 (+18\%) & 146 (+21\%) & 5967 (-35\%) & 49 (-31\%) & 44 (-39\%) & 14844\\
\textbf{Trim$_{L}$ + \kleemath$_{R}$} & \textbf{149 (+18\%)} & \textbf{152 (+26\%)} & \textbf{5699 (-38\%)} & \textbf{48 (-32\%)} & \textbf{40 (-44\%)} & \textbf{14850}\\
\end{tabular}
}
\label{tab:klee}
\end{table*}

Since programs usually have infinitely many execution paths, it is necessary
to enforce some resource bounds when running \klee. In particular, we run
\klee with a timeout of 300 seconds and a limit of 64 on the number of forks
(i.e., symbolic branches).

The results of our evaluation are presented in Table~\ref{tab:klee}.  Here,
the column labeled {\sc Safe} shows the number of programs for which \klee
explores all execution paths without reporting any errors or
warnings.\footnote{By warning, we mean any internal \klee warning that
  designates an incompleteness in \klee's execution (e.g., solver timeouts and
  concretizing symbolic values).} Hence, these programs can be considered
verified. The second column, labeled {\sc Unsafe}, shows the number of
programs reported as buggy by each variant of \klee. In this context, a bug
corresponds to an explicit assertion violation in the program. Next, the third
column, labeled {\sc Paths}, shows the number of program paths that \klee
explored for each variant. Note that fewer paths is better---this means that
\klee needs to explore fewer executions before it finds the bug or proves the
absence of an assertion violation. The next two columns measure the number of
programs for which each \klee variant reaches a resource limit. In particular,
the column labeled {\sc Timeout} shows the number of programs for which \klee
fails to terminate within the 5-minute time limit. Similarly, the column {\sc
  Max-Forks} indicates the number of programs for which each \klee variant
reaches the limit that we impose on the number of forks.  Finally, the last
column, labeled {\sc Time}, shows the total running time of each \klee variant
on all benchmarks.
%

As shown in Table~\ref{tab:klee}, program trimming increases the number
of programs that can be proved safe by 16--18\%. Furthermore, program trimming
allows \klee to find up to 30\% more bugs within the given resource limit. In
addition, \klee with program trimming needs to explore significantly fewer
paths (up to 38\%) and reaches the resource bound on significantly fewer
programs. Finally, observe that the overall running time of \klee decreases by
up to 30\%.

\begin{figure}[b!]
\vspace{-1em}
\scalebox{1}{
\begin{tikzpicture}
\pgfkeys{/pgf/number format/set thousands separator = {\textrm{'}}}
\begin{semilogyaxis}[
    compat=1.9,
    log ticks with fixed point, 
    table/col sep=tab, 
    unbounded coords=jump, 
    filter discard warning=false, 
    /pgfplots/table/y index=3,
    /pgfplots/table/header=false,
    tick label style={font=\footnotesize},
    legend style={font=\footnotesize},
    label style={font=\footnotesize},
    legend cell align=left,
    axis y line*=none,
    axis x line*=bottom,
    xlabel=\# of solved benchmarks,
    ylabel=Time (s),
    xmin=0,
    ymin=0.01,
    ymax=200,
    xtick={0,50,100,150,200,250,300,350},
    mark size=1pt,
    width=8cm,
    height=5.3cm,
    legend entries={\kleemath$_{\mathit{BFS}}$, Trim$_{L+B}$ + \kleemath$_{\mathit{BFS}}$, Trim$_{L}$ + \kleemath$_{\mathit{BFS}}$},
    every axis legend/.append style={at={(0,1)}, anchor=north west, outer xsep=5pt, outer ysep=5pt,},
  ]
  \addplot+[Black,mark=o,line join=round,mark options={scale=1.5,fill=none},mark repeat=20,semithick] table {evaluation/logs/klee/bfs/klee-baseline/klee-notrim-no-ms.quantile.csv};
  \addplot+[CornflowerBlue,mark=triangle,line join=round,mark options={scale=1.5,fill=none},mark repeat=20,semithick] table {evaluation/logs/klee/bfs/klee-trimming-loops-bounded/klee-trim-loops-bounded.quantile.csv};
  \addplot+[NavyBlue,mark=square,line join=round,mark options={scale=1.5,fill=none},mark repeat=20,semithick] table {evaluation/logs/klee/bfs/klee-trimming-loops/klee-trim-loops.quantile.csv};
\end{semilogyaxis}
\end{tikzpicture}}
\vspace{-0.1in}
\caption{Quantile plot of time and solved benchmarks for selected \klee variants.}
\label{fig:time}
\end{figure}
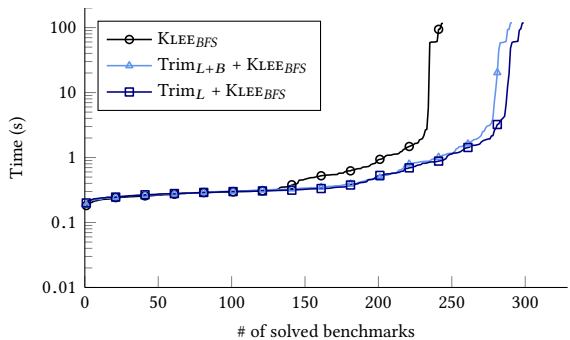

Figure~\ref{fig:time} compares the number of benchmarks solved by the original
version of \klee (using BFS) with its variants using program
trimming. Specifically, the x-axis shows how many benchmarks were solved
(i.e., identified as safe or unsafe) by each variant (sorted by running
time), and the y-axis shows the corresponding running time per benchmark. For
instance, we can see that \textbf{Trim$_{L}$ + \kleemath$_{\mathit{BFS}}$}
  solves 246 benchmarks within less than one second each, whereas the original
  version of \klee only solves 203 benchmarks.

\emph{Summary.}
Overall, the results shown in Table~\ref{tab:klee} and Figure~\ref{fig:time}
demonstrate that program trimming significantly improves the
effectiveness and performance of a mature, state-of-the-art symbolic execution
tool. In particular, program trimming allows \klee to find more bugs and prove
more programs correct within a given resource limit independently of its
search strategy.

\subsection{Threats to Validity}

We identified these threats to the validity of our experiments:
\begin{itemize}[leftmargin=1.1em]
\item \emph{Sample size}: We used 439 programs, most of which, however, are
  taken from the SV-COMP benchmarks, a widely-used and established set of verification
  tasks.
\item \emph{Safety checkers}: We evaluate our technique using two safety
  checkers, which, however, are mature and representative of two program
  analysis techniques.
\item \emph{Trimming configurations}: We only presented experiments using a
  selection of the different configurations that \toolname provides (see
  Section~\ref{sect:implementation}). However, all of these configurations
  are orthogonal to each other, and we evaluated a large variety of them
  to demonstrate the benefits of our technique.
\end{itemize}

\section{Related Work}
\label{sect:relatedWork}

The ideas in this paper are related to a long line of previous work
on program transformations and safety precondition inference.

\emph{Program slicing.}
One of the most well-known program simplification techniques is \emph{program
  slicing}, which removes program statements that are not relevant to some
criterion of interest (e.g., value of a variable at some program
point)~\cite{Weiser1981,AgrawalHorgan1990,Tip1995,BinkleyGallagher1996}. A
program slice can be computed either statically or dynamically and includes
both forward and backward variants. Program trimming differs from traditional
program slicing in two ways: first, trimming focuses on removing execution
paths as opposed to statements; second, it is meant as a pre-processing
technique for safety checkers rather than a transformation to aid program
understanding.
\mv{In particular, a typical slicing tool may not produce compilable and
  runnable code that could be consumed by subsequent safety checkers.}

More semantic variants of program slicing have also been considered in later
work~\cite{BallNaik2003,JhalaMajumdar2005,FieldRamalingam1995,ComuzziHart1996,CanforaCimitile1998,HarmanHierons2001}.
For instance, Jhala and Majumdar propose \emph{path slicing} to improve the
scalability of software model checkers~\cite{JhalaMajumdar2005}. In
particular, path slicing eliminates all operations that are irrelevant toward
the reachability of the target location in a given program path. Unlike
program trimming, path slicing is not used as a pre-processing step and works
on a single program path that corresponds to a counterexample trace.

Prior work has also considered how to slice the program with respect to a
predicate~\cite{FieldRamalingam1995,ComuzziHart1996,CanforaCimitile1998,HarmanHierons2001}.
Such techniques can be useful for program understanding, for example, when the
user only wants to see statements that affect a given condition (e.g., the
predicate of a conditional). In contrast, program trimming is not meant as a
program understanding technique and removes program paths that are irrelevant
for a given safety property. Furthermore, the trimmed program is not meant for
human consumption, as it semantically prunes program paths through the
insertion of \code{assume} statements.

\mv{In general, slicing has been used before invoking a program
  analyzer~\cite{JaffarMurali2014,MillettTeitelbaum2000,HatcliffDwyer2000,DolbyVaziri2007,IvancicYang2005,ChebaroKosmatov2012,ChoiPark2015}. A
  key difference with these approaches is that the result of trimming is valid
  code, which compiles and runs, instead of an abstract representation, such
  as a control flow graph or model.}

\emph{Pre-processing for program analyzers.}
In the same spirit as this paper, prior work has also used program
transformations to improve the precision or scalability of program
analyzers~\cite{GurfinkelWei2008,LalQadeer2014-MixedSemantics,SharmaDillig2011,ChristakisMueller2016,Christakis2015,Wuestholz2015,ChristakisWuestholz2016}. For
instance, a transformation for faster goal-directed
search~\cite{LalQadeer2014-MixedSemantics} moves all assertions to a single
main procedure with the goal of speeding up analysis.  Another program
transformation called \emph{loop splitting} aims to improve the precision of
program analyzers by turning \emph{multi-phase} loops into a sequence of
\emph{single-phase} loops~\cite{SharmaDillig2011}. However, neither of these
techniques instrument the program with assumptions to guide safety checking
tools.

Recent techniques rely on the \mv{verification} results of a \mv{full-fledged}
analyzer, such as an abstract interpreter or a model checker, to guide
automatic test case generation
tools~\cite{ChristakisMueller2016,Christakis2015,DacaGupta2016,CzechJakobs2015} or other
static analyzers~\cite{ChristakisMueller2012,Wuestholz2015,ChristakisWuestholz2016,BeyerHenzinger2012}, \mv{some
  even using slicing as an intermediate step~\cite{CzechJakobs2015}}. \mv{In
  contrast, program trimming is more lightweight by not relying on previous
  analyzers and, thus, can be used as a pre-processing step for any safety
  checker.}

\emph{Precondition inference.}
%
The use of precondition inference dates back to the dawn of
program verification~\cite{Dijkstra1975}. Most verification techniques infer a
sufficient condition for program safety and prove the correctness of the
program by showing the validity of this
condition~\cite{Hoare1969,Dijkstra1975,Hoare1971,HoareHe1987,FlanaganLeino2002,BarnettChang2005,BarnettLeino2005,Moy2008,ChandraFink2009}. In
this work, we do not aim to infer the weakest
possible safety precondition; instead, we use lightweight, modular static
analysis to infer \emph{a} sufficient condition for safety. Furthermore, we
use safety conditions to prune program paths rather than to verify the
program.

Program trimming hinges on the observation that the negation of a sufficient
condition for property $P$ yields a necessary condition for the negation of
$P$. Prior program analysis techniques also exploit the same
observation~\cite{DilligDillig2008,DilligDillig2010,DilligDillig2011,ZhangNaik2013}. For
instance, this duality has been used to perform modular path-sensitive
analysis~\cite{DilligDillig2008} and strong updates on elements of unbounded
data structures~\cite{DilligDillig2010,DilligDillig2011}.

While most program analysis techniques focus on the inference of sufficient
preconditions to guarantee safety, some techniques also infer \emph{necessary
  preconditions}~\cite{LogozzoLahiri2014,LogozzoBall2012,CousotCousot2013,CousotCousot2011,NaikYang2012}.
For example, Verification Modulo Versions (VMV) infers both necessary and
sufficient conditions and utilizes previous versions of the program to reduce
the number of warnings reported by
verifiers~\cite{LogozzoLahiri2014}. Similarly, necessary conditions are
inferred to repair the program in such a way that the repair does not remove
any ``good'' traces~\cite{LogozzoBall2012}. Finally, the techniques described
by Cousot et al. infer necessary preconditions, which are used to improve the
effectiveness of the Code Contracts abstract
interpreter~\cite{CousotCousot2013,CousotCousot2011,FahndrichLogozzo2010}.

\emph{Abductive reasoning.} There has been significant work on program
analysis using abductive reasoning, which looks for a \emph{sufficient}
condition that implies a desired
goal~\cite{LiDillig2013,AlbarghouthiDillig2016,ZhuDillig2013,CalcagnoDistefano2009,DilligDillig2013-Inference,DilligDillig2013-Explain,DilligDillig2014}. Our
analysis for computing safety conditions can be viewed as a form of abductive
reasoning in that we generate sufficient conditions that are stronger than
necessary for ensuring safety. However, we perform this kind of reasoning in a
very lightweight way without calling an SMT solver or invoking a logical
decision procedure.

\emph{Modular interprocedural analysis.}
The safety condition inference we have proposed in this paper is modular in
the sense that it analyzes each procedure independently of its callers. There
are many previous techniques for performing modular (summary-based)
analysis~\cite{PnueliSharir1981,CalcagnoDistefano2009,DilligDillig2008,AikenBugrara2007,YorshYahav2008}.
Our technique differs from these approaches in several ways: First, our
procedure summaries only contain safety preconditions, but not
post-conditions, as we handle procedure side effects in a very conservative
way. Second, we do not perform fixed-point computations and achieve soundness
by initializing summaries to $\mathit{false}$. Finally, we use
summary-based analysis for program transformation rather than verification.

\emph{Property-directed program analysis.}
There is a significant body of work that aims to make program analyzers
property directed. Many of these techniques, such as
BLAST~\cite{HenzingerJhala2002,HenzingerJhala2004,BeyerHenzinger2007},
SLAM~\cite{BallMajumdar2001,BallRajamani2001,BallRajamani2002}, and
YOGI~\cite{GodefroidNori2010,NoriRajamani2009} rely on
counterexample-guided abstraction refinement (CEGAR)~\cite{ClarkeGrumberg2000}
to iteratively refine an analysis based on counterexample traces.  Another
example of a property-directed analysis is the IC3/PDR
algorithm~\cite{Bradley2011,HoderBjorner2012}, which iteratively performs
forward and backward analysis for bounded program executions to decide
reachability queries. Although abstract interpretation is traditionally not
property directed, there is recent work~\cite{RinetzkyShoham2016} on adapting
and rephrasing IC3/PDR in the framework of abstract interpretation. In
contrast, we propose a general pre-processing technique
to make any eager program analysis property directed.

\emph{Path-exploration strategies.}
Most symbolic execution and testing techniques utilize different strategies
to explore the possible execution paths of a program. For example, there are
strategies that prioritize ``deeper paths'' (in depth-first search),
``less-traveled paths''~\cite{LiSu2013}, ``number of new instructions
covered'' (in breadth-first search), ``distance from a target
line''~\cite{MaKhoo2011}, or ``paths specified by the
programmer''~\cite{SenTanno2015}. In the context of symbolic execution,
program trimming can be viewed as a search strategy that prunes safe paths and
steers exploration toward paths that are more likely to contain bugs. However,
as shown in our experiments, our technique is beneficial
independently of a particular search strategy.

\section{Conclusion}
\label{sect:conclusion}

In this paper, we have proposed \emph{program trimming}, a program
simplification technique that aims to reduce the number of execution paths
while preserving safety. Program trimming can allow any safety checker to be goal directed
by pruning execution paths that cannot possibly result in an assertion violation. Furthermore, because
our proposed trimming algorithm is very lightweight, it can be used as an effective pre-processing
tool for many program analyzers. As shown by our evaluation, program trimming allows an
abstract interpreter, namely \crab, to verify 21\% more programs while cutting running
time by 30\%. Trimming also allows \klee, a dynamic symbolic execution engine, to find more bugs and verify more
programs within a given resource limit.



In future work, we plan to investigate the impact of program trimming on other
kinds of program analyzers, such as bounded model checkers. We also plan
to investigate alternative program trimming algorithms and strategies.

\section*{Acknowledgments}
We would like to thank Cristian Cadar and Martin Nowack for their help with
\klee. We would also like to thank the main developer of \crab, Jorge Navas,
for his help with \crab and for his positive feedback on the usefulness of
program trimming.
\mv{We are grateful to Microsoft for partly funding the first three authors
  and to the anonymous reviewers for their constructive feedback. This work is
  supported by AFRL Award FA8750-15-2-0096 and NSF Award \#1453386. The views and conclusions contained herein are those of the authors and should not be interpreted as
necessarily representing the official policies or endorsements, either expressed or implied, of DARPA or
the U.S. Government.}

\newpage

\bibliographystyle{ACM-Reference-Format}
\bibliography{tandem}

\ifdefined\EXTENDEDVERSION
\appendix
\newpage
\section{Proof of Theorem~4.6}

\begin{proofsketch}
For most statements (e.g., assignment, assumption, assertion), $\tc'$ is just
the standard weakest  precondition of $s$ with respect to $\tc$.

For heap reads and writes, we already argued why $\tc' \Rightarrow
\mathit{wp}(s, \tc)$.
The heap allocation rule is also correct since it ``havocs'' the allocated
pointer.

The correctness of the procedure call rule follows from the following two
facts: First, $\mathit{summary}(\mathit{prc}, \summary, \bar{v})$ is a
conservative safety condition for the call to
$f$. In particular, if $f \in \mathit{dom}(\summary)$, this follows from the
soundness of $\summary$.  If $f \not \in \mathit{dom}(\summary)$,
$\mathit{false}$ (resp. $\mathit{true}$) is a sufficient condition for the
safety of any procedure that does (resp. does not) contain an
assertion. Second, we ``havoc'' the value of any memory location modified in
$f$. The correctness of our $\mathit{havoc}$ operation follows from (a) the
correctness of the $\mathit{store}$ function, and (b) $\forall v. \phi
\Rightarrow \mathit{wp}(v:= e, \phi)$ for any expression $e$.
\end{proofsketch}

\section{Proof of Theorem~5.1}

\begin{proof}
The proof is by induction on the number of statements (i.e., $n-i$).

Suppose $i=n$. If $s_n$ is not an assertion, then the safety condition is
$\mathit{true}$, so we add \code{assume} $\mathit{false}$. Since $s_n$ can never fail,
$\mathit{false}$ is indeed necessary for failure. If $s_n$ is
\code{assert} $\phi$, then the necessary condition for failure is $\neg
\phi$. Since the safety condition for $s_n$ is $\phi$, our technique
instruments the code with \code{assume} $\neg \phi$.

For the inductive step, suppose $i < n$ and let:
\[ \aliasing, \summary, \mathit{true} \vdash s_{i+1}; \ldots; s_n: \tc \]
By the inductive hypothesis, $\neg \tc$ is a necessary condition for the
failure of $s_{i+1}, \ldots, s_n$. We consider three cases: (1)~$s_i$ is
an assertion \code{assert} $\phi$.  Then, the necessary condition for the
failure of $s_i; \ldots; s_n$ is $\neg \phi \lor \neg \tc$. Since the safety
condition for $s_i; \ldots; s_n$ is $\phi \land \tc$, our technique
instruments the code with \code{assume} $\neg \phi \lor \neg \tc$.  (2)~If
$s_i$ is an assumption \code{assume} $\phi$, the necessary condition for
failure is $\phi \land \neg \tc$, which is exactly the trimming condition
computed by our technique. (3)~Otherwise, the necessary condition for failure
is $\mathit{wp}(s_i, \neg \tc)$. Suppose $\aliasing, \summary, \tc \vdash s_i:
\tc'$. By soundness of the safety condition inference, we have $\tc'
\Rightarrow \mathit{wp}(s_i, \tc)$, and we instrument the code with
\code{assume} $\neg \tc'$. Since $s_i$ is neither an assertion nor an
assumption, we have $\mathit{wp}(s_i, \neg \tc) \equiv \neg \mathit{wp}(s_i,
\tc) $; thus, $\mathit{wp}(s_i, \neg \tc) \Rightarrow \neg \Phi'$.
\end{proof}

\fi

\end{document}